\documentclass[12pt]{iopart}

\usepackage{graphicx}
\usepackage{bm}
\usepackage{braket}
\usepackage{times}
\expandafter\let\csname equation*\endcsname\relax
\expandafter\let\csname endequation*\endcsname\relax
\usepackage{amsmath}
\usepackage{amssymb}
\usepackage{amstext}
\usepackage[sort&compress,numbers]{natbib}

\renewcommand{\c}{\hat c}
\newcommand{\cd}{\hat c^\dagger}
\newcommand{\Erec}{E_\mathrm{rec}}
\newcommand{\BI}{\mathrm{BI}}

\renewcommand{\L}{\mathrm{L}}
\newcommand{\R}{\mathrm{R}}
\newcommand{\BIket}{\ket{\mathrm{BI}}}
\newcommand{\BIbra}{\bra{\mathrm{BI}}}

\begin{document}

\title{Large-Amplitude Superexchange of High-Spin Fermions in Optical Lattices} 

\author{Ole J\"urgensen, Jannes Heinze, Dirk-S\"oren L\"uhmann}
\address{Institut f\"ur Laser-Physik, Universit\"at Hamburg, Luruper Chaussee 149, 22761 Hamburg, Germany
}
 

\begin{abstract}
We show that fermionic high-spin systems with spin-changing collisions allow to monitor superexchange processes in optical superlattices with large amplitudes and strong spin fluctuations. By investigating the non-equilibrium dynamics, we find a superexchange dominated regime at weak interactions. The underlying mechanism is driven by an emerging tunneling-energy gap in shallow few-well potentials. As a consequence, the interaction-energy gap that is expected to occur only for strong interactions in deep lattices is reestablished.
By tuning the optical lattice depth, a crossover  between two regimes with negligible particle number fluctuations is found: first, the common regime with vanishing spin-fluctuations in deep lattices and, second, a novel regime with strong spin fluctuations in shallow lattices. We discuss the possible experimental realization with ultracold $^{40}$K atoms and observable quantities in double wells and two-dimensional plaquettes. 
\end{abstract}

\pacs{37.10.Jk, 3.75.Ss, 71.10.Fd, 75.30.Et}

\maketitle


\section{Introduction} \label{sec_introduction}
Spin-exchange interactions are of fundamental importance for the magnetic properties of strongly correlated media \cite{Heisenberg1926, Dirac1926, Heisenberg1928, Dirac1929}. Of particular interest are superexchange interactions that are induced by higher-order tunneling processes via virtually occupied intermediate states \cite{Kramers1934, Anderson1950}. In solid state systems such as CuO and MnO, theses processes are of long-ranged nature, as they do not depend on a direct wave-function overlap of the electrons in contrast to direct exchange interactions. In the context of high-temperature superconductivity they are in the focus of current investigations \cite{Lee2006}. Experiments with ultracold atoms in optical lattices present a promising tool for the study of fundamental properties of higher-order tunneling processes \cite{Lewenstein2007}. Here, typically second-order superexchange tunneling is considered. However, also models with third-order tunneling processes as the leading order are 
discussed in the literature, e.g., strongly correlated fermions in frustrated lattices \cite{Pollmann2006}. Recently, the direct time-resolved observation of superexchange processes has been demonstrated in experiments with ultracold atoms in deep double-well potentials \cite{Trotzky2008}. These experiments exploit the interaction blockade inhibiting first-order tunneling processes. However, for the interaction blockade the spin fluctuations arising from superexchange are diminished due to the small amplitude. There is great current interest in accomplishing long-ranged anti-ferromagnetic ordering with ultracold fermions in optical lattices. However, the small energy associated with superexchange cannot compete with thermal fluctuations in present experiments with isotropic lattices \cite{Greif2013}.

The realization of fermionic high-spin systems with ultracold atoms promises a deeper understanding of fundamental spin-spin interactions \cite{Taie2010, Stellmer2011, Lompe2010, Krauser2012}. These systems are of broad interest, since they offer the possibility to realize, e.g., a multitude of new quantum phases \cite{Rodriguez2010, Lecheminant2005, Wu2005, Rapp2007, Ho1999, Wu2003, Tu2006}, the simulation of SU(N) magnetism \cite{Honerkamp2004, Hermele2009, Taie2010, Gorshkov2010} as well as spin changing collisions \cite{Ho1998, Law1998, Ohmi1998, Bornemann2008}. Recent experimental advances demonstrate long-lived coherent spin dynamics of fermionic $^{40}K$ atoms in optical lattices with tunable effective spin and reliable initial-state preparation \cite{Krauser2012}. The interplay of spin-changing collisions and tunneling processes leads to new phenomena such as the instability of an initially band insulating state \cite{Krauser2012}. This allows for the detailed study of many-body non-equilibrium 
dynamics in a highly controllable environment.

Here, we investigate the non-equilibrium superexchange dynamics of fermionic high-spin systems with spin-changing collisions in optical double-well and plaquette lattices. We find a regime dominated by superexchange with large amplitude and strong spin fluctuations in \textit{shallow} potentials. Usually, the observation of superexchange with ultracold quantum gases requires deep lattices with a large interaction gap $U$ that suppresses the first-order tunneling $J$. In the latter regime, the superexchange has small amplitudes $J^2/U$ and is therefore extremely sensitive to thermal fluctuations. In a high-spin system, spin-changing collisions lift the Pauli blocking of an initially prepared band-insulator state with $m_f=\pm 1/2$. This causes the initial state to be unstable which in general allows strong particle-number fluctuations to arise \cite{Krauser2012}. Surprisingly, we find that the first-order tunneling is strongly suppressed in shallow few-well potentials. In this regime, an emerging tunneling-
energy gap leads to an interaction blockade at large $J/U$. We show how this phenomenon can be used to investigate superexchange processes with large amplitude corresponding to short time scales. This opens new possibilities for the dynamical study of superexchange in a non-equilibrium system, previously restricted to deep optical lattices with small superexchange amplitude $J^2/U$.

We present a comprehensive study of the exact time evolution of a band insulator in double-well potentials and four-well plaquettes. A possible experimental realization with $^{40}K$ and a measurement scheme for particle-number fluctuations is discussed. The required techniques, i.e., the preparation of optical few-well potentials \cite{Sebby-Strabley2006, Sebby-Strabley2007, Anderlini2007, Trotzky2008, Nascimbene2012} and the reliable initial state preparation \cite{Krauser2012} have already been successfully demonstrated.

\section{Superexchange in Spin-3/2 Systems} \label{sec_spin-3/2}
\begin{figure}
\centering
 \includegraphics[width=0.50\textwidth]{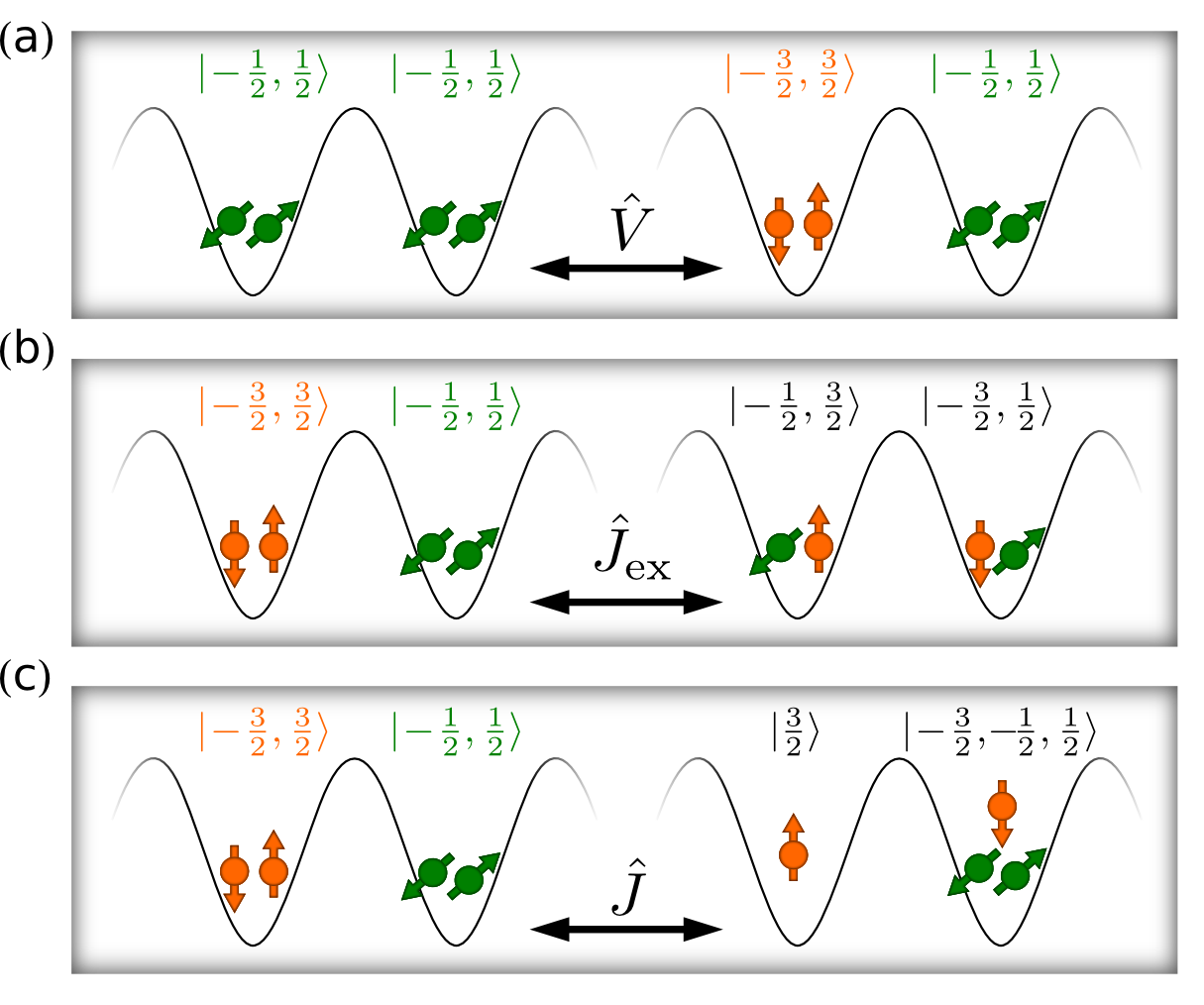}\caption{(a) In a spin $f=\frac32$ system, a spin-changing collision transfers particles on a lattice site from $m_f=\pm\frac12$ to $m_f=\pm\frac32$ and vice versa. (b) A superexchange process exchanges two particles on neighboring lattice sites. Here the orange $m_f=-\frac32$ particle on the left site is exchanged with the green $m_f=-\frac12$ particle on the right site. (c) A first-order tunneling processes inducing particle-number fluctuations. \label{fig_schemes}} 
\end{figure}

\begin{figure}
\centering
 \includegraphics[width=0.9\textwidth]{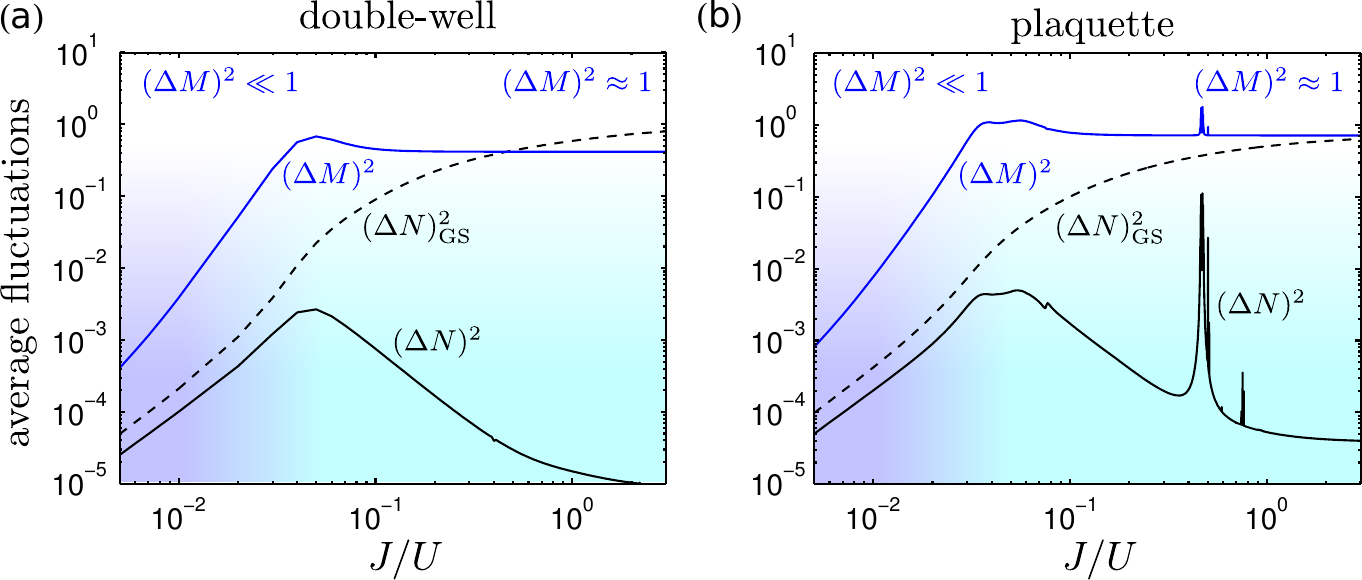}\caption{(a) The solid black line shows the particle-number fluctuations $(\Delta N)^2$ and the blue line the spin fluctuations $(\Delta M)^2$ averaged over a long period of time ($t<5000\, h/U$) as a function of $J/U$. The dashed lines shows the fluctuations $(\Delta N)^2_\mathrm{GS}$ of the respective ground-state of the system for comparison. The shading illustrates the crossover between the two regimes $(\Delta M)^2 \ll 1$ and $(\Delta M)^2 \approx 1$. (b) The same data evaluated for the time evolution in a four-well plaquette. Additional to the two limiting regimes, sharp resonances with high fluctuations occur for specific values of $J/U$.
\label{fig_fluctuations}} 
\end{figure}

In this paper we discuss the realization of a pure superexchange model with high-spin fermions. First, we introduce the corresponding superexchange as well as the full microscopic Hamiltonian. In the subsequent sections, we demonstrate that the superexchange Hamiltonian
\begin{equation}\begin{split}
\label{eq:seH}
  \hat H_\mathrm{ex} = &V \sum_i \cd_{i, -\frac32} \cd_{i, \frac32} \c_{i, -\frac12} \c_{i, \frac12} + c.c. \\
+ &\frac{J^2}{U} \sum_{\braket{i,j}} \sum_{\sigma,\sigma'} \cd_{i,\sigma} \c_{i,\sigma'} \cd_{j,\sigma'} \c_{j,\sigma} + c.c.
\end{split}\end{equation}
can be realized with particles of spin $f=\frac32$ in an optical lattice in a surprisingly wide range of parameters, including shallow lattices with large amplitudes $J^2/U$. Here, $\cd_{i, \sigma}$ ($\c_{i, \sigma}$) creates (annihilates) a particle on lattice site $i$ in the spin-state $\sigma \in \{-\frac{3}{2}, -\frac{1}{2}, \frac{1}{2}, \frac{3}{2}\}$ and obeys the fermionic commutation relations. The first part of this Hamiltonian describes spin-changing collisions with an amplitude $V$ as depicted in figure \ref{fig_schemes}(a). Two particles on a lattice site exchange their spins while preserving the total magnetization, i.e., $\ket{-\frac12,\frac12} \leftrightarrow \ket{-\frac32,\frac32}$. 
The second part of the Hamiltonian \eqref{eq:seH} represents superexchange processes (figure \ref{fig_schemes}(b)). These second-order tunneling processes exchange particles on neighboring lattices sites, which preserves the density distribution but can lead to strong spin-fluctuations already in a double-well system. The corresponding amplitude $J^2/U$ accounts for the increased interaction energy $U$ in the virtually occupied intermediate state. The interaction energy itself is omitted in \eqref{eq:seH} as a constant offset, since neither spin-changing collisions nor the superexchange affects the particle-number distribution. The first-order tunneling (figure \ref{fig_schemes}(c)) is not present in the Hamiltonian \eqref{eq:seH} and any intersite dynamics is mediated by superexchange processes.

A system of fermionic particles with spin $f=\frac32$ in an optical lattice can be used to realize the Hamiltonian \eqref{eq:seH}. The full microscopic description is given by the Hamiltonian
\begin{equation}\begin{split}
\label{eq:h}
 \hat H_\mathrm{full} =\ &\frac{U}{2} \sum_i \hat n_i (\hat n_i - 1) \\
- &J \sum_{\braket{i,j}} \sum_\sigma \cd_{i, \sigma} \c_{j, \sigma} + c.c. \\
+ &V \sum_i \cd_{i, -\frac32} \cd_{i, \frac32} \c_{i, -\frac12} \c_{i, \frac12} + c.c.,
\end{split}\end{equation}
The number operator $\hat n_i$ counts particles on a lattice site $i$. In contrast to the superexchange Hamiltonian, the full model allows for first-order tunneling $\hat J$, e.g., $\ket{-\frac32,\frac32}\ket{-\frac12,\frac12} \rightarrow \ket{\frac32}\ket{-\frac32,\frac12,\frac12}$ (figure \ref{fig_schemes}(c)) and a corresponding change in onsite-interaction energy $U$. The spin-changing collisions represent a small perturbation of the standard Hubbard Hamiltonian $\hat H_\mathrm{HM} = \hat U + \hat J$. We assume an amplitude of $V=0.01\, U$, which is a realistic value for $^{40}K$ (see section \ref{sec_spin-dependent}). 

In this work, we investigate the dynamics of the band insulating ground state
\begin{equation}
\label{eq:BI}
 \BIket = \prod_i \hat f^\dagger_{i, -\frac{1}{2}} f^\dagger_{i, \frac{1}{2}} \ket{0}.
\end{equation}
of the unperturbed Hamiltonian  $\hat H_\mathrm{HM}$. The robust experimental preparation of this two-component band insulator as well as the control over the spin-changing collisions $\hat V$ via a magnetic field has been demonstrated in reference \cite{Krauser2012}. In the following we discuss the dynamics of \eqref{eq:BI} in a double-well and a four-well plaquette potential. Both setups can be realized experimentally \cite{Sebby-Strabley2006, Sebby-Strabley2007, Anderlini2007, Trotzky2008, Nascimbene2012} and can be solved numerically exact. We show that the time evolution is governed by the superexchange model \eqref{eq:seH} even in the absence of a direct interaction blockade. This effect depends on the existence of the fluctuationless eigenstate \eqref{eq:BI} of the unperturbed Hamiltonian in shallow lattices. Note that there is no such eigenstate in bosonic systems where the realization of a pure superexchange model is only possible in deep lattices.

\section{Double-well} \label{sec_double-well}

By superimposing two optical lattices of a short wavelength $\lambda_\mathrm{short}$ and a longer wavelength $\lambda_\mathrm{long}=2\ \lambda_\mathrm{short}$ an array of double-well potentials can be realized \cite{Sebby-Strabley2006, Sebby-Strabley2007, Anderlini2007, Trotzky2008}. The preparation of the initial state \eqref{eq:BI} can be achieved by loading atoms with $m_f=\pm 1/2$ into the optical lattice of the short wavelength before ramping up the second lattice to separate the individual double-wells.

The time evolution can be determined by means of exact diagonalization of the microscopic Hamiltonian \eqref{eq:h}. For a wide range of parameters $J/U$ we investigate the dynamics and evaluate the time-dependent particle-number fluctuations $(\Delta N)^2 = \braket{\hat n_i^2}-\braket{\hat n_i}^2$. In figure \ref{fig_fluctuations}(a) the time-averaged value of $(\Delta N)^2$ is shown as a function of $J/U$ (solid black line) and compared with the corresponding value of the ground state (dashed line). For deep lattices, fluctuations are exponentially suppressed for decreasing $J/U$ due to an interaction blockade $U$. Surprisingly and in contrast to the ground state, we also find an exponential suppression of $(\Delta N)^2$ in shallow lattices for increasing $J/U$. Therefore, the time evolution of the band insulator is governed by the superexchange Hamiltonian \eqref{eq:seH} for almost all lattice depths. This phenomenon allows the simulation of the superexchange Hamiltonian with ultracold atoms without an 
interaction blockade, i.e., with large amplitudes $J^2/U$.

\begin{figure*}
\centering
 \includegraphics[width=0.9\textwidth]{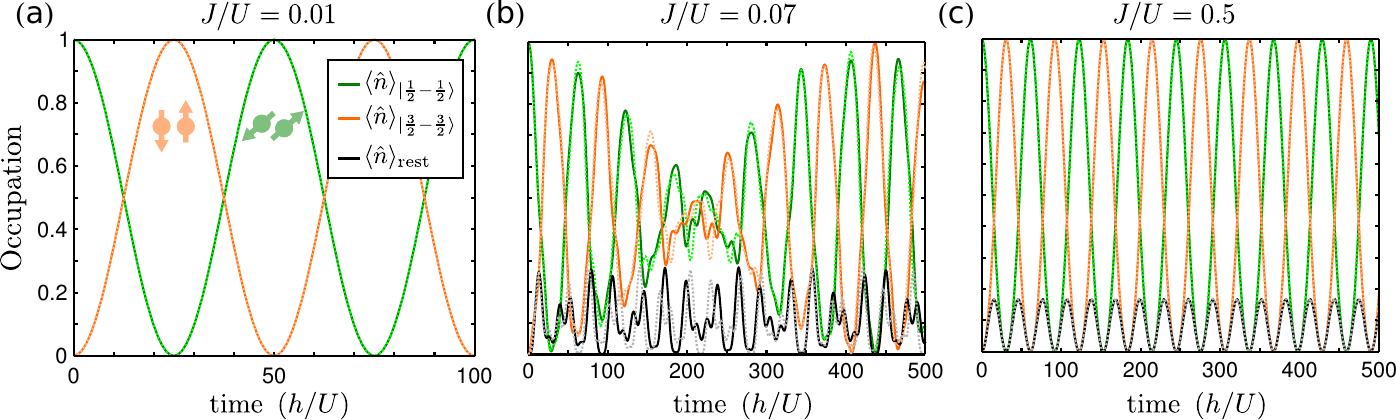}\caption{Time-evolution of the band-insulator \eqref{eq:BI} in a double-well. Shown is the probability to find the state $\ket{-\frac12, \frac12}$ (green lines) or $\ket{-\frac32, \frac32}$ (orange line) in one well. The black lines correspond to any other configuration. Note that due to symmetry reasons the population is identical in both wells. The solid lines correspond to the full Hamiltonian \eqref{eq:h} and the dotted lines to superexchange Hamiltonian \eqref{eq:seH}. For (a) and (c) both models lead to the same results. \label{fig_double-well}} 
\end{figure*}

In the case of vanishing tunneling $J \ll U$, i.e., in a very deep lattice, both sites undergo in-phase Rabi oscillations induced by the spin-changing collisions (figure \ref{fig_double-well}(a)). The state remains fluctuationless due to a large interaction blockade $U$. After an evolution time of $t=h/4V$, the state will eventually be the band-insulator $\tilde\BIket = \ket{-\frac32,\frac32}\ket{-\frac32,\frac32}$. In both states, $\BIket$ and $\ket{\tilde \BI}$, no tunneling is possible due to the Pauli principle. For all other times, however, the state is in a superposition of $\ket{-\frac12,\frac12}$ and $\ket{-\frac32,\frac32}$ on both wells. For a finite amplitude $J \neq 0$ first-order tunneling is in principle possible (Figure \ref{fig_schemes}(c)). Surprisingly, the numerical results show very low occurrence of particle-number fluctuations $(\Delta N)^2$ even and especially for large values of $J/U$. Thus, the dynamics of \eqref{eq:BI} is described by the superexchange operator \eqref{eq:seH}. 
Figure \ref{fig_double-well} shows the time evolution for various parameters $J/U$ calculated with the full Hamiltonian \eqref{eq:h} (solid lines) and with the superexchange Hamiltonian \eqref{eq:seH} (dotted lines). Only at values of $J/U\approx 0.05$, where $(\Delta N)^2$ has a maximum, finite deviations between the two models are present (figure \ref{fig_double-well}(b)). In shallow lattices, periodic oscillations with full amplitude are reestablished (figure \ref{fig_double-well}(c)). Superexchange processes take place on a short time-scale, leading to the population of onsite states with finite magnetization $M_i = \sum_\sigma n_{i,\sigma}\sigma$ such as $\ket{\frac12 \frac32}$ (black lines in figure \ref{fig_double-well}(b,c)). This directly corresponds to a non-zero value of spin fluctuations $(\Delta M)^2 = \braket{\hat M_i^2}-\braket{\hat M_i}^2$, whereas particle-number fluctuations $(\Delta N)^2$ are negligible. In contrast, in deep lattices spin fluctuations remain small. Here, $(\Delta M)^2$ is on the order of $10^{-3}$ and thus the black line in figure \ref{fig_double-well}(a) is not visible. The time-averaged value of the spin fluctuations in the evolution of \eqref{eq:BI} is shown as a blue line in figure \ref{fig_fluctuations}(a). We can identify two distinct regimes: one with almost no spin fluctuations $\Delta M \ll 1$ for small $J/U$ and one with large spin fluctuations $\Delta M \approx 1$ for large $J/U$.

\section{Plaquette} \label{sec_plaquette}

\begin{figure*}
 \includegraphics[width=1\linewidth]{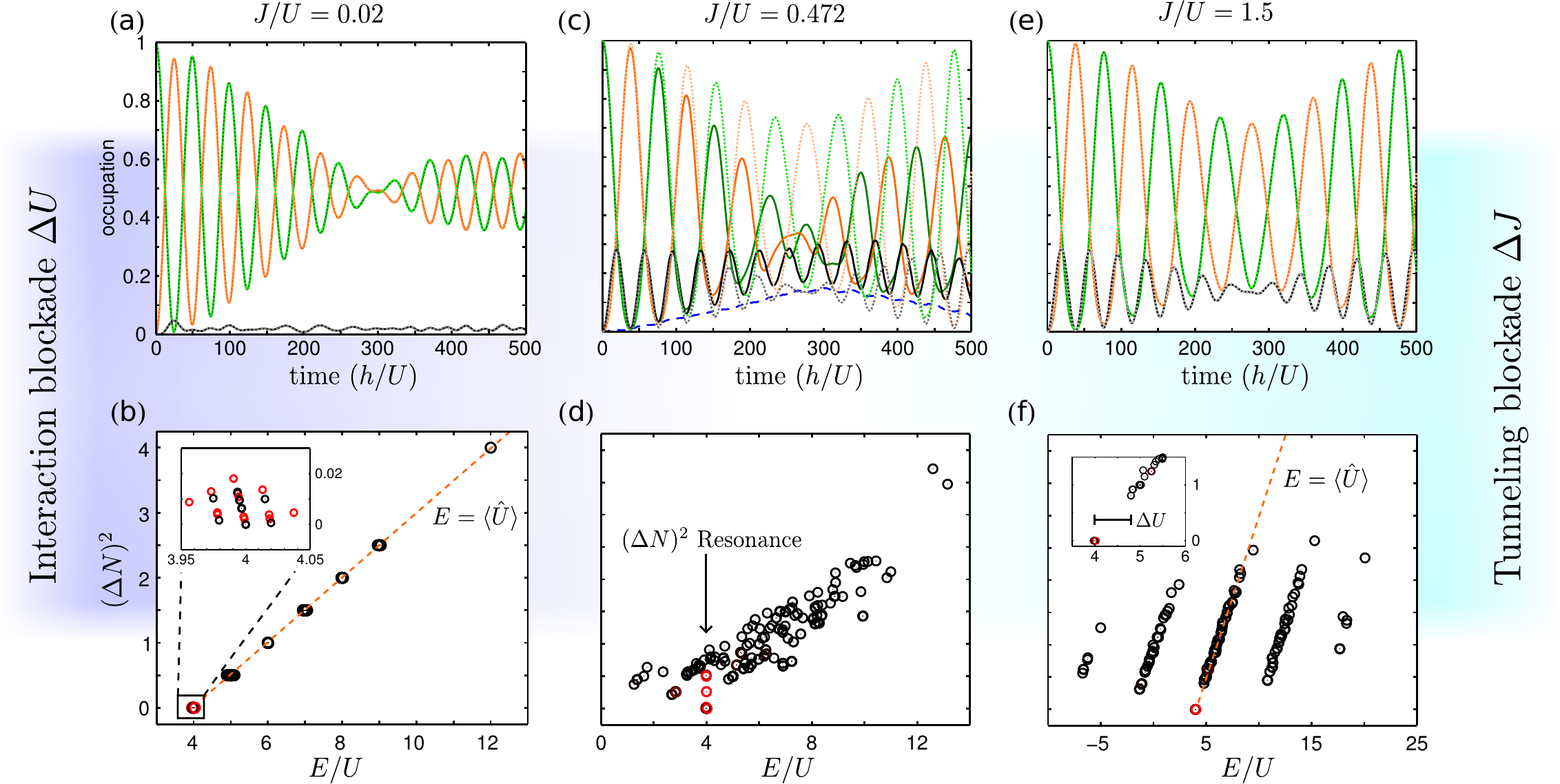}\caption{(a,c,e) Time evolution of an initially prepared band-insulator in a four-well plaquette for various values of $J/U$. The green lines correspond to sites with $\ket{-\frac12,\frac12}$, the orange lines to sites with $\ket{-\frac32,\frac32}$. The black lines correspond to all other configurations that can only be populated by first-order tunneling or superexchange processes. The dotted lines show the time evolution using the superexchange Hamiltonian \eqref{eq:seH} that gives the same results for (a) and (e). The dotted blue line in (c) corresponds to fluctuations $(\Delta N)^2$, which cause the superexchange model to fail. (b,d,f) Spectra depicting eigenstates $\ket{\phi}_i$ with energy $E_i$ and fluctuations $(\Delta N)^2$. Every marker represents an eigenstate, where the red color indicates finite overlap with the initial state \eqref{eq:BI}. In deep lattices (b) and for eigenstates with $\braket{\hat J}=0$ in shallow lattices (f) the 
eigenenergies are given by the interaction energy $E=\braket{\hat U}$. \label{fig_plaquette}}
\end{figure*}
 
The crossover between the two distinct regimes discussed above is not a unique feature of the double-well but occurs also in other few-well systems. In the following we identify the underlying mechanism of the fluctuationless time-evolution in shallow potentials for the particular case of a four-well plaquette. This mechanism is, however, general for all few-well systems that show the discussed behavior.

 A four-well plaquette is the natural extension of a double-well potential in two dimensions. It is equivalent to a one-dimensional four-well lattice with periodic boundary conditions and has been recently realized experimentally \cite{Nascimbene2012}. Despite the increased complexity compared with the double-well, the Hamiltonian \eqref{eq:h} can still be diagonalized for a four-well lattice allowing to investigate the time-evolution of the initially prepared band-insulator exactly. The Hamiltonian has a number of symmetries that can be exploited to reduce the numerical effort. These symmetries are the spatial translation, spin-flip, quadrupole and particle-hole symmetry. Furthermore, the initial state \eqref{eq:BI} also obeys several symmetries and we can restrict the basis to the corresponding subspace.

We investigate a wide range of parameters and find the two regimes with $(\Delta N)^2 \approx 0$, known from the double-well. The spin fluctuations $(\Delta M)^2$ that arise from the initial state \eqref{eq:BI} and the ground state properties of the system (figure \ref{fig_fluctuations}(b)) are very similar to the double-well system. In contrast to the results of the double-well, however, strong fluctuations appear at certain values of $J/U$. The eigenspectra allow for a qualitative understanding of the two limits $J \ll U$ and $J \gg U$ and also explain the emerging resonances with high fluctuations $(\Delta N)^2$.

In deep potentials $J \ll U$, the onsite interaction $U$ is the dominating energy scale of the Hamiltonian and the eigenstates separate in groups of Fock states with integer $U$-excitations that are perturbed by $\hat J$ and $\hat V$. The spectrum is shown in Figure \ref{fig_plaquette}(b), where each marker corresponds to an eigenstate $\ket{\phi_i}$ with an energy $E_i$ and particle-number fluctuations $(\Delta N)^2$. The latter directly correspond to the interaction energy $\braket{\hat U}$
\begin{equation} \label{eq:dNofU}
(\Delta N)^2 = 2 \frac{\braket{\hat U} - \BIbra \hat U \BIket}{U}.
\end{equation}
Red markers have finite overlap with the initial state, which is clearly confined to the group of eigenstates without fluctuations. Here, the interaction blockade $\Delta U$ causes the dynamics to be determined by the superexchange Hamiltonian \eqref{eq:seH}, which is used in conventional realizations with ultracold atoms \cite{Trotzky2008}. The time-evolution of the initial state \eqref{eq:BI} is shown in Figure \ref{fig_plaquette}(a), where the solid lines correspond to the full model \eqref{eq:h} and the dotted lines to the superexchange model \eqref{eq:seH}. Note that due to the very small amplitude of the superexchange $J^2/U$ spin fluctuations $(\Delta M)^2$ are suppressed and merely cause a slow dephasing of the onsite Rabi oscillations.

In very shallow potentials, the tunneling operator $\hat J$ is the dominating energy scale of the Hamiltonian \eqref{eq:h}. The spectrum for $J/U=1.5$ is shown in figure \ref{fig_plaquette}(f). In a finite lattice, the spectrum is gaped due to the discrete energies of the Bloch functions. In the case of four lattice sites and taking into account the conservation of momentum, the gap is $\Delta J = 4J$. When this is large compared with $U$ and $V$, the spectrum separates in groups with $\braket{\hat J} = z\, \Delta J$ where $z$ is an integer number. The initial state \eqref{eq:BI} obeys $\hat J \BIket = 0$ and $\hat U \BIket = 4U \BIket$ and consequently lies in the group with $\braket{\hat J}=0$. For $V=0$, in this group the eigenenergies are given by the interaction energy $E_i=\braket{\hat U}_i$, similar to the regime of the interaction blockade, as indicated by the dashed lines in the spectra. 
Since the spin-changing collisions present only a small perturbation $V \ll U$, the band isolator state \eqref{eq:BI} is only perturbed by states with $E_i \approx 4U$. For these states, the particle-number fluctuations \eqref{eq:dNofU} are small and the band isolator state can only develop small particle-number fluctuations. Interestingly, a subordinate energy gap $\Delta U \approx U$ arises between fluctuationless states with $E_i \approx 4U$ (including $\BIket$) and states with particle-number fluctuations and an energy $E_i > 4U+\Delta U$. Due to the tunneling-energy gap, the interaction blockade is reestablished in sufficiently shallow potentials, where now $V$ is the competing energy scale instead of $J$. The gap $\Delta U$ amplifies the suppression of particle-number fluctuations for the initial state, but is not a necessary requirement for $(\Delta N)^2 \ll 1$. The surprisingly low fluctuations in very shallow potentials thus arise from a tunneling-energy gap $\Delta J$ and a subordinate interaction blockade. In figure \ref{fig_plaquette}(e) the time evolution for $J/U=1.5$ is shown, where no deviations between the full model \eqref{eq:h} and the superexchange model \eqref{eq:seH} can be seen. The time evolution of figure \ref{fig_plaquette}(e) in units of $h/U$ is identical for all lattice depths with $J/U > 0.2$ and sufficiently small fluctuations $(\Delta N)^2$, i.e., away from resonances in figure \ref{fig_fluctuations}(b).

In the parameter regime $0.2 < J/U < 1.5$, neither an interaction blockade $\Delta U$ nor a tunneling-energy blockade $\Delta J$ inhibits particle-number fluctuations (see figure \ref{fig_plaquette}(d)). The eigenenergies are not determined by the interaction energy only and thus do not directly correspond to the particle-number fluctuations. However, few eigenstates have finite overlap with the initial state \eqref{eq:BI} and thus, only at specific parameters $J/U$ fluctuations arise. At these resonances, the time evolution of the full model and the superexchange model differ significantly, as can be seen in figure \ref{fig_plaquette}(c). The superexchange model (dotted line) predicts the same behavior as for very shallow potentials, whereas the correct calculations are strongly affected by the occurrence of triply occupied sites, which is plotted as a dashed blue line.

\section{Measuring particle-number fluctuations} \label{sec_measureN} 
\begin{figure}
\centering
 \includegraphics[width=0.65\textwidth]{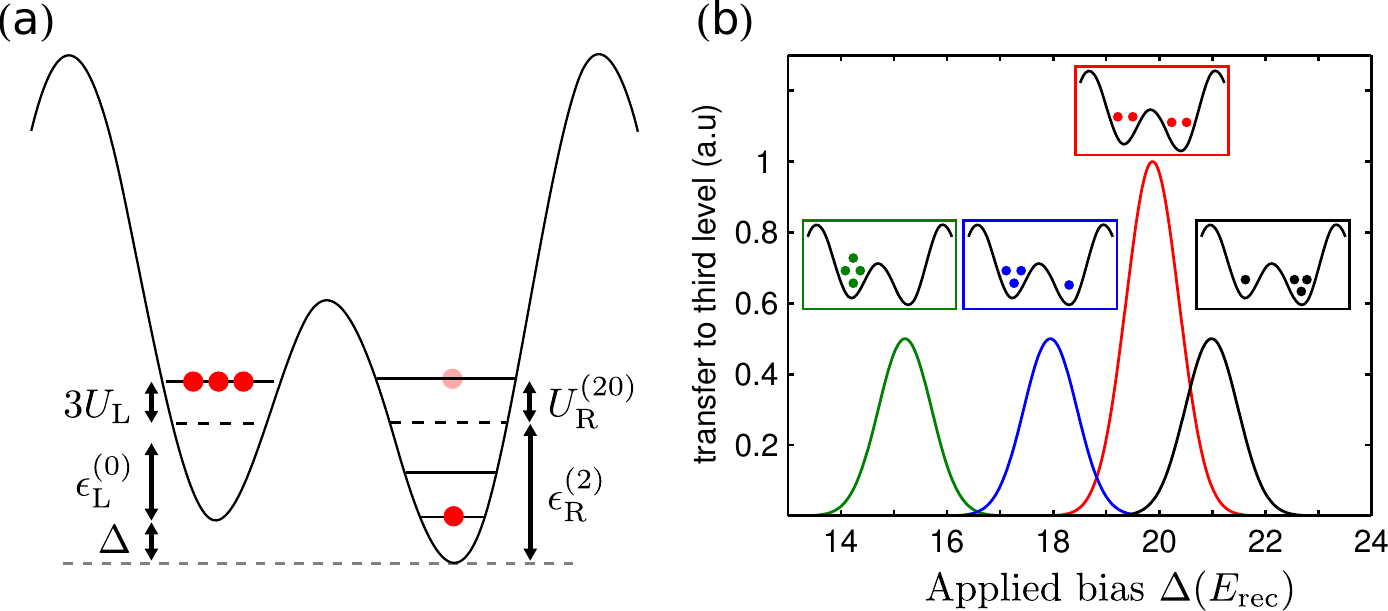}\caption{(a) Scheme to measure triply occupied sites. The lowest mode of the left well can be tuned into resonance with the third mode of the right well by applying a bias $\Delta\ $, where the interaction-energy shifts have to be taken into account. (b) The bias can be tuned to fulfill the resonance condition for certain particle-number distributions $n_\L|n_\R$, allowing for the measurement of triply occupied wells (blue line). \label{fig_measure_triplons}} 
\end{figure}
An experimental proof of superexchange processes and the absence of first-order tunneling is the demonstration of strong spin fluctuations and vanishing particle-number fluctuations. The former can be accessed in an experiment with $^{40}K$ via the population of new spin components, as discussed in section \ref{sec_spin-dependent}. The latter can be measured in an optical double-well setup with the technique illustrated in figure \ref{fig_measure_triplons}(a). First, the lattice with the short wavelength $\lambda_\mathrm{short}$ is ramped up to freeze the populations. Afterwards, the relative phase between the two lattice beams is tuned to introduce a bias $\Delta$ between the wells. That way the third vibrational mode of the deeper well can be brought into resonance with the first one of the shallower well. The required bias $\Delta=\epsilon_\R^{(2)}-\epsilon_\L^{(0)} - (n_\L-1) U_\L + U^{(2 0)}_\R n_\R$ between the left (L) and the right (R) well depends on the single-particle energies $\epsilon^{(i)}$ of 
the vibrational modes $i$ and the particle-number distribution $n_\L|n_\R$ that introduces interaction energy shifts. When a tunneling process occurs, the interaction energy on the left well is reduced by $(n_\L-1) U_\L$, while it is increased on the right well by $U^{(2 0)}_\R n_\R$, where $U^{(2 0)}$ is given by the wave-function overlap of the lowest mode with the third one. This technique has already been successfully demonstrated in reference \cite{Cheinet2008}. For every given particle-number distribution we schematically  plot the resonance of a transfer to the third band as a function of the applied bias $\Delta$ in figure \ref{fig_measure_triplons}(b). The blue line corresponds to the case of three particles in the left and one in the right well. We assume a width of $0.5 \Erec \gg J$ for the resonances, which is a pessimistic assumption compared with the measured width in the experiment of reference \cite{Cheinet2008}. Figure \ref{fig_measure_triplons}(b) shows that it is clearly possible to tune the bias to a value such that particles will be transferred to the third band only if there are exactly three particles on the shallower site. The population of the third mode can be measured by means of the band-mapping technique afterwards.

\section{Spin-dependent interaction energies} \label{sec_spin-dependent}

\begin{figure*}
 \includegraphics[width=1\linewidth]{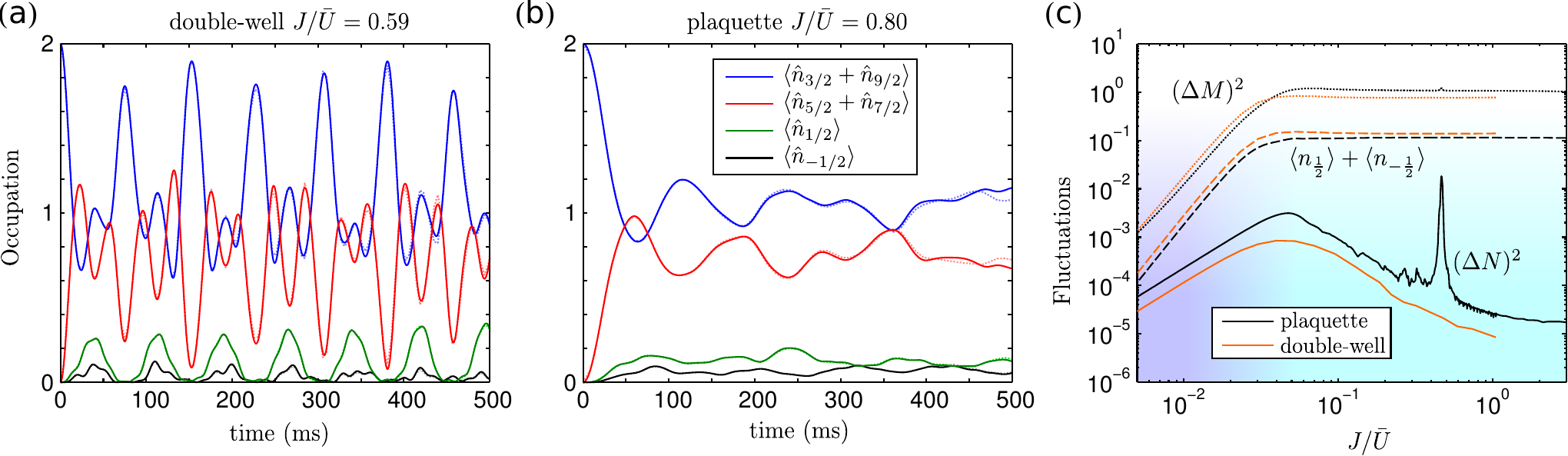}\caption{(a,b) Time-evolution with 6 components and $^{40}K$ parameters in the double-well and the four-well plaquette. The blue lines denote particles in the initial spin-states $m_f=\frac32$ and $m_f=\frac92$ and the red line corresponds to the spin-states $m_f=\frac52$ and $m_f=\frac72$. The black and green lines depict the population of the spin-states $m_f=-\frac12$ and $m_f=\frac12$, respectively, that can only be populated consecutive to tunneling/superexchange processes. The dotted lines represent the time evolution in the superexchange model \eqref{eq:seH}. The time scale is given by the absolute energies obtained from Wannier functions of a lattice with lattice constant $a = 515\, \mathrm{nm}$ and a transversal lattice depth of $V_\perp = 25\, \Erec \ (35\, \Erec)$ in two (one) directions for the double-well (plaquette) and a lattice depth $V_\parallel$ that is tuned to adjust $J/\bar U$. (c) Average particle-number fluctuations $(\Delta N)^2$ and spin fluctuations $(\Delta M)^2$ (dotted lines) as a function of $J/\bar{U}$. The dashed lines show the average number of particles in the spin components $m_f=-\frac12$ and $m_f=\frac12$. The behavior is qualitatively identical to the $f=\frac32$ system. Note that the time average is taken over an evolution time of up to $t_\mathrm{max} = h/2 \bar V$, corresponding to $40-350\, \mathrm{ms}$ depending on the lattice depth. \label{fig_6_components}}
\end{figure*}


As a concrete example, we now consider $^{40}K$ prepared in the $f=9/2$ manifold as a possible species for an experimental realization of high-amplitude superexchange. As the initial state we suggest a mixture of $m_f=9/2$ and $m_f=3/2$, which allows for rich spin-dynamics and is the most similar to a band insulator in a $f=3/2$ system. Without tunneling, i.e., in very deep lattices, Rabi-oscillations between $\ket{\frac32 \frac92}$ and $\ket{\frac52 \frac72}$ occur in phase on every lattice site in analogy to the $f=3/2$ system. When tunneling is allowed, however, more complicated dynamics arise. Consecutive to tunneling processes, new spin-components can be populated by other spin-changing collisions, e.g. $\ket{\frac32 \frac72}\rightarrow\ket{\frac12 \frac92}$. States with $m_f < -1/2$ are only occupied by third-order processes and we neglect them in our simulations as their influence on the qualitative behavior is assumed to be small. This reduces the system effectively to 6 spin components. The spin 
fluctuations $(\Delta M)^2$ are no directly measurable quantity, however the population of the new spin components is proof for the occurrence of tunneling or superexchange processes.

Another difference compared with the spin-3/2 system is that the interaction energies $U$ and $V$ are no longer spin-independent but slightly differ from each other \cite{Krauser2012}. The generalization of the microscopic Hamiltonian \eqref{eq:h} to 6 spin components is straight forward, when the spin-dependent interaction energies are known. For our simulations we use parameters from \cite{Krauser2012}. As a characteristic quantity for the potential depth we use the ratio $J/\bar U$, where the interaction energy $\bar U$ corresponds to a spin-preserving collision in the initial state. The energy $\bar V$ of the spin-changing collision $\ket{\frac32 \frac92} \leftrightarrow \ket{\frac52 \frac72}$ is significantly smaller with $\bar V/\bar U \approx 0.007$. The quadratic Zeeman-energy must be used to tune the spin-changing collisions into resonance, because the initial and final state of a spin-changing collision have different interaction energies. Of course, this cannot be done for all spin states at the 
same time and the exact time-behavior depends on the applied magnetic field. Our simulations are performed with $B=169\,\mathrm{mG}$, but any other value near the resonance is suitable. 

In figure \ref{fig_6_components}(a,b) the time-behavior of the $m_f$-components is plotted for the double-well and the plaquette. They differ from the behavior of a $f=3/2$ system with spin-independent interaction energies in several ways. In the case of spin-dependent interactions the superexchange operator $\hat J_\mathrm{ex}$ can be approximated by the matrix elements 
\begin{equation}
 \braket{f|\hat J_\mathrm{ex}|i} = \sum_k \frac{\braket{f|\hat J|k} \braket{k|\hat J|i}}{\braket{\hat U}_k - \braket{\hat U}_i/2 - \braket{\hat U}_f/2},
\end{equation}
between an initial state $\ket{i}$ and a final state $\ket{f}$ with exactly two particles per lattice site, respectively. The sum is over all possible intermediate states  $\ket{k}$ and $\braket{\hat U}_i=\braket{i|\hat U|i}$. For flat lattices, the time-evolution can be described by this operator as can be seen by comparing the dotted lines with the solid lines in figure \ref{fig_6_components}(a,b). The new spin-components $m_f=1/2$ and $m_f=-1/2$ occur after a few milliseconds, which corresponds to an increase in spin fluctuations $(\Delta M)^2$. In the plaquette the coherent Rabi-oscillations are damped out rapidly in strong contrast to the spin-3/2 system, where they last for arbitrarily long times. The reason for this is the larger number of different frequencies due to the spin-dependency of the interaction strengths.

Concerning the appearance of particle-number fluctuations $(\Delta N)^2$, however, the qualitative behavior is identical to the spin-3/2 system, as shown in figure \ref{fig_6_components}(c). The population of the components $m_f=-\frac12$ and $m_f=\frac12$ (dashed lines) is proportional to the average spin fluctuations $(\Delta M)^2$ (dotted lines). This is an intriguing advantage of the high spin $f=\frac92$ as opposed to a pure spin-3/2 system. Only for certain values of $J/U$ resonances with triply occupied sites occur, while for a wide regime of parameters, the state remains fluctuationless. Due to the additional components, spin fluctuations increase to $\Delta M \approx 1$ even in deep lattices for very long evolution times. The particle-number fluctuations, as discussed in section \ref{sec_measureN}, as well as the population of the new spin-components are quantities that are accessible experimentally. In combination they are direct evidence for superexchange processes in the absence of direct 
tunneling. In conclusion, $^{40}K$ is a suitable atomic species for the proposed experiments to simulate superexchange Hamiltonians in shallow lattices.

\section{Conclusions}
\label{sec_conclusions}

We have found a mechanism for fermionic high-spin systems that allows for the investigation of superexchange processes in shallow lattices, i.e., at large amplitudes $J^2/U$. This regime corresponds to much faster time scales than in deep lattices and was not accessible yet. The non-equilibrium dynamics of an initially band-insulating state with two components has been calculated, and we found the intersite dynamics to be fully dominated by the superexchange. Although, in general, spin-changing collisions circumvent Pauli blocking and therefore allow for strong particle-number fluctuations in shallow lattices, we have shown that the first-order tunneling is strongly suppressed in few-well potentials while strong spin fluctuations arise via superexchange processes. The underlying mechanism is driven by an emerging tunneling-energy gap that reestablishes an interaction gap, which is usually only expected in deep lattices. Furthermore, we have investigated the crossover of superexchange regimes 
with weak and strong spin fluctuations in deep and shallow lattices, respectively. 
 
We have discussed the experimentally relevant case of optical double-well and plaquette potentials. We have simulated the time evolution for a pure spin-3/2 system as well as for realistic parameters of $^{40}K$ and have shown that in both cases it is governed by the superexchange Hamiltonian. The latter allows for a measurement of spin fluctuations via the population of certain spin components. The double-well potential and the four-well plaquette are both suited for an experimental realization.  Even larger lattices of up to six sites in one spatial direction show the existence of a sufficiently large tunneling-energy gap $\Delta J \approx J$ and a subordinate interaction-energy blockade. The necessary experimental tools, i.e., the optical potentials for a double-well in one and two dimensions  \cite{Sebby-Strabley2006, Sebby-Strabley2007, Anderlini2007, Trotzky2008, Nascimbene2012} as well as the preparation of the initial band insulator with $^{40}K$ atoms have both been demonstrated successfully \cite{Krauser2012} . 

\section{Acknowledgments}
We would like to thank K.~Sengstock, L.~Mathey and J.S.~Krauser for very fruitful discussions. We thank the Deutsche Forschungsgesellschaft (DPG) for financial support within grant FOR 801 and grant GRK 1355.


\end{document}